\documentclass[useAMS,usenatbib]{mn2e}

\def\aap{AA}
\def\apjl{ApJL}

\def\mnras{MNRAS}
\def\apj{ApJ}
\def\aj{AJ}

\def\nat{Nat}
\def\araa{ARAA}
\usepackage{graphicx}
\usepackage{float}
\usepackage{bm} 
\usepackage{amssymb}
\usepackage{amsmath} 
\usepackage[export]{adjustbox}
\usepackage{mathrsfs} 
\usepackage{mathtools}

\usepackage{color}

\title[Stream gaps from GMCs]{Gaps in globular cluster streams: giant molecular clouds can cause them too}
\author[]{Nicola C. Amorisco$^{1,2}$\thanks{E-mail:
nicola.amorisco@cfa.harvard.edu}, Facundo A. G{\' o}mez$^{2}$, Simona Vegetti$^{2}$, Simon D.M. White$^{2}$  \\
$^{1}$Institute for Theory and Computation, Harvard-Smithsonian Center for Astrophysics, 60 Garden St., MS-51, Cambridge, MA 02138, USA\\
$^{2}$Max Planck Institute for Astrophysics, Karl-Schwarzschild-Strasse 1, D-85740 Garching, Germany
}

\begin{document}



\maketitle

\label{firstpage}

\begin{abstract}
As a result of their internal dynamical coherence, thin stellar
streams formed by disrupting globular clusters (GCs) can
act as detectors of dark matter (DM) substructure in the Galactic
halo.  Perturbations induced by close flybys amplify into detectable
density gaps, providing a probe both of the abundance and of the
masses of DM subhaloes.  Here, we use N-body simulations to show that
the Galactic population of giant molecular clouds (GMCs) can also
produce gaps (and clumps) in GC streams, and so may confuse the
detection of DM subhaloes. We explore the cases of streams analogous
to the observed Palomar 5 and GD1 systems, quantifying the expected
incidence of structure caused by GMC perturbations. Deep observations
should detect such disturbances regardless of the substructure content
of the Milky Way's halo. Detailed modelling will be needed to
demonstrate that any detected gaps or clumps were produced by DM
subhaloes rather than by molecular clouds.
\end{abstract}


\begin{figure*}
\centering
\includegraphics[width=\textwidth]{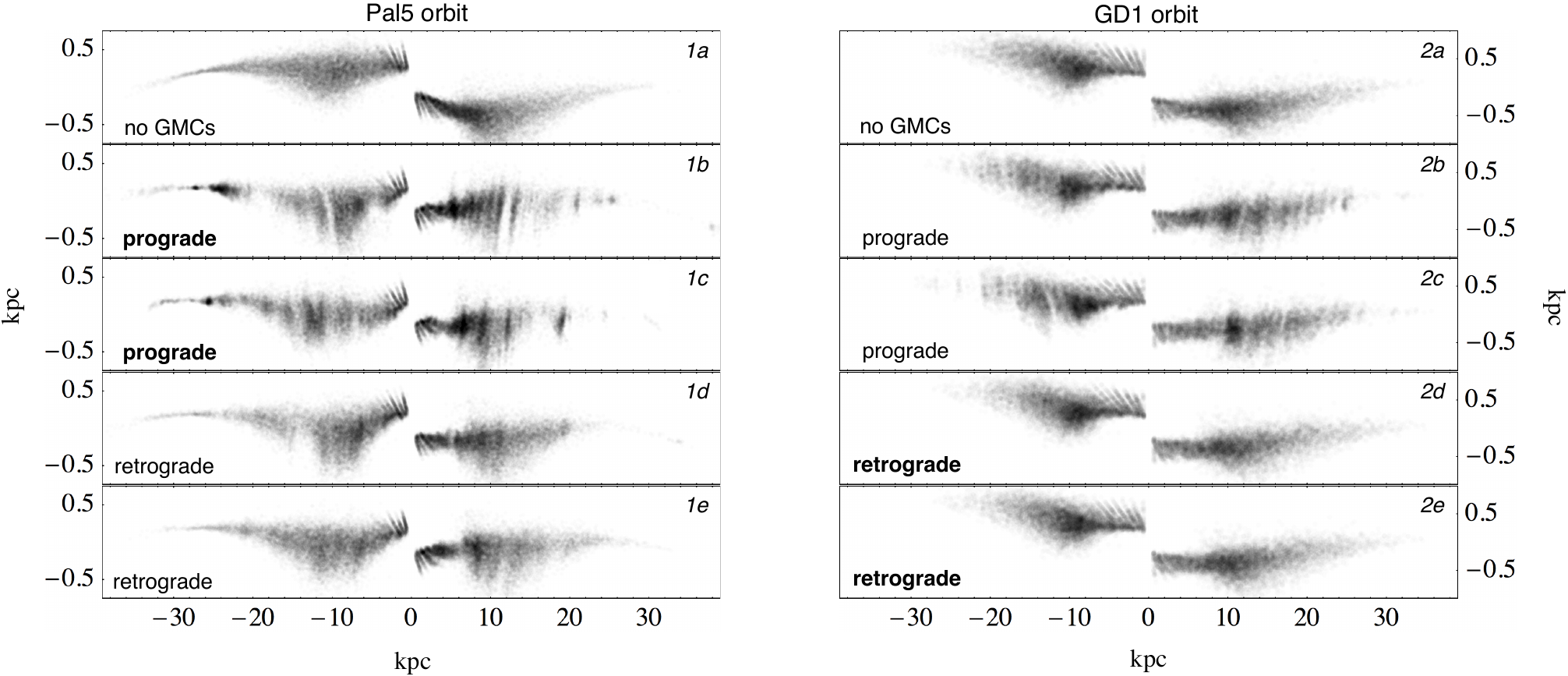}
\caption{Examples of GC streams on Pal5- and GD1-like orbits,
  straightened along the remnant's trajectory.  Panels {\it 1a} and
  {\it 2a} show unperturbed streams. All other panels show the stream
  when a disk of giant molecular clouds is added. Panels {\it 1b-c}
  and {\it 2b-c} show streams on prograde orbits.  Panels {\it 1d-e} and
  {\it 2d-e} show the corresponding retrograde streams, where the
  increased relative velocity of close encounters reduces perturbation
  amplitudes. 
  }
\label{fig2}
\end{figure*}

\section{Introduction}
 
The ability to reproduce large-scale
cosmological observations
is a great success of the standard $\Lambda$CDM structure formation
model \citep[e.g.,][]{PL14,GT05,AN14} but is independent of the nature
of the dark matter.  Various dark matter models share this success,
but make different predictions for the amount and properties of
structure on scales of galaxy haloes and smaller. Standard WIMP/axion 
CDM 
predicts galaxy haloes that host a multitude
of bound subhaloes, the remnants of earlier halo generations
\citep[DSs, e.g.,][]{M99,K99,G04,D08,VS08}. Such a population is
substantially suppressed in warm DM models as a result of a cut-off in
the linear power spectrum at small mass scales \citep[$M\lesssim 10^8
  M_\odot$, e.g.,][]{CAV00,BOT01,ML12,AM13}, while self-interacting DM
models 
also predict both a reduction in the DS population
and variations in their internal structural properties
\citep[e.g.,][]{MV12,Ro13}. The characterisation of low mass DSs thus
represents a promising route to advance our understanding of DM.


Great interest has been sparked by the realisation that thin globular
cluster (GC) streams in the Milky Way (MW) could reveal the
presence of DSs, by showing distinctive signs of flyby encounters
\citep{RI02,KJ02}. The internal dynamics of kinematically cold streams
are such that perturbations induced by a gravitational encounter are
amplified with time: coherent shifts in the energy distribution of
stream stars near closest approach result in the formation of
under-dense regions, usually referred to as
gaps \citep[e.g.,][]{Yo11,RC13,DE15a,JS16} as well as over-dense clumps
and even kinks \citep[e.g.][]{Ng14,Ng16}.  This makes GC streams with
internal velocity dispersions of a few km/s potentially
sensitive to encounters of DSs with masses as small as $M\sim10^5
M_\odot$. The detection of a population of such DSs would put
stringent constraints on the nature of DM.

Two spectacular examples of thin GC streams are those of Palomar 5
\citep[Pal5,][]{Od02, Od03, GD06a} and GD1 \citep{GD06b}, for both of
which the density profile can be mapped over $\sim$10 kpc in length
using SDSS data \citep[e.g.,][]{RC12,SK10}. Interestingly, both
streams display substructure with amplitude that appears marginally
higher than expected due to observational (counting) noise alone
\citep{RC12,RC16}, suggesting the possibility that external
perturbations are indeed important.

In this Letter, we show that the Galactic population of giant molecular clouds (GMCs) is
expected to produce gaps and clumps in the Pal5 and GD1 streams which are similar
to those that would be induced by DSs, thus significantly complicating the detection
of halo substructure.
We explore this process using N-body simulations of stream formation 
in the presence of a disk of GMCs, but ignoring any DM substructure.
We quantify the abundance and properties of the gaps and clumps
produced, thus characterising the confusing background against which
the effects of DSs must be indentified.


\section{N-body simulations}

We use Gadget2 \citep{VS05} to compute the evolution of GC streams
with orbits similar to Pal5 and GD1 within a three-component, static
approximation to the MW's potential taken from \cite{RE15}.  In all
runs, the progenitor GC is represented by a self-gravitating,
isotropic Plummer sphere with a mass of $10^{4.8} M_\odot$, a
half-mass radius of $r_{\rm h}=16$ pc, and $10^5$ equal-mass
``stars''.  In agreement with the inferred orbital properties
\citep[e.g.][]{Od02, GD06a, AK15, TF15}, our Pal5-like runs have $(r_{\rm
  per}, r_{\rm apo}, i) = (8~{\rm kpc}, 19~{\rm kpc}, 65^\circ)$,
where $r_{\rm per}$, $r_{\rm apo}$ and $i$ are respectively
pericentric radius, apocentric radius and inclination with respect to
the stellar disk. This results in an orbital period
$T_r=0.30$~Gyr. Our GD1-like runs have
$(r_{\rm per}, r_{\rm apo}, i) = (12~{\rm kpc}, 27~{\rm kpc},
35^\circ)$ \citep[e.g.][]{GD06b,Wi09, SK10, AB15} for a period of
$T_r=0.44$~Gyr.  All runs start with the GC at its orbital apocenter,
and follow it for 10 Gyr.  While Pal5 and GD1 are observed to be
prograde and retrograde, respectively, with respect to the sense of
rotation of the stellar disk, we explore both prograde and retrograde
configurations for both orbits.

Our GMC populations have a total mass of $M_{\rm tot} = 10^9 M_\odot$, 
69\% of which lie between a radius of 2~kpc and the solar radius, $R_{\odot}=8.5$~kpc,
the remainder is at larger radii.
GMCs are confined to a razor-thin disk that, at $R_{\odot}$, breaks from an inner
scale-length of 2.5~kpc to an outer one of 3.9~kpc \citep{HD15}. We
generate individual GMC masses by adopting the results of \citet{Ri16}, which report
a steep GMC mass function outside $R_{\odot}$ (an average power-law slope of 
$\alpha\sim-2.2$ for masses $M_{GMC}>10^5~M_{\odot}$, where 
$dN/d M_{GMC}\propto M_{GMC}^{-\alpha}$). The mass function within $R_{\odot}$
is less steep ($\alpha\sim-1.5$), although it truncates sharply with no GMCs at
$M_{GMC}\geq10^7~M_{\odot}$. {This model implies
19$\pm3.6$ GMCs with masses $M_{GMC}>10^6~M_{\odot}$ and orbiting at $R>8$~kpc, 
which could interact with the Pal5 stream; 6.9$\pm$2.7 of these could interact with GD1 too, 
since they orbit at $R>12$~kpc.}
Mass and radius distributions are generated using different
random seeds in different runs, and we explore 15 such initial
conditions for each orbit. In each run, GMCs are represented by
smoothed massive particles, using the gravitational softening kernel
adopted by Gadget2 \citep[][]{VS05}.  {Conservatively}, we use a softening length
of $\epsilon = 100$ pc, independent of GMC mass, corresponding to
the size of the largest Galactic GMCs \citep[e.g.,][]{Ev99,La05}. 
{ We fix the maximum time step to 0.07 Myr, so that, at a relative velocity of 300 km/s, 
the crossing of a length $\epsilon$ is resolved with a minimum of $\sim$4.5 steps}. GMC
lifetimes are much shorter than the dynamical timescales involved
here, but we assume the population remains stationary and enforce this
by letting our GMCs live through the full duration of each run.

\begin{figure*}
\centering
\includegraphics[width=\textwidth]{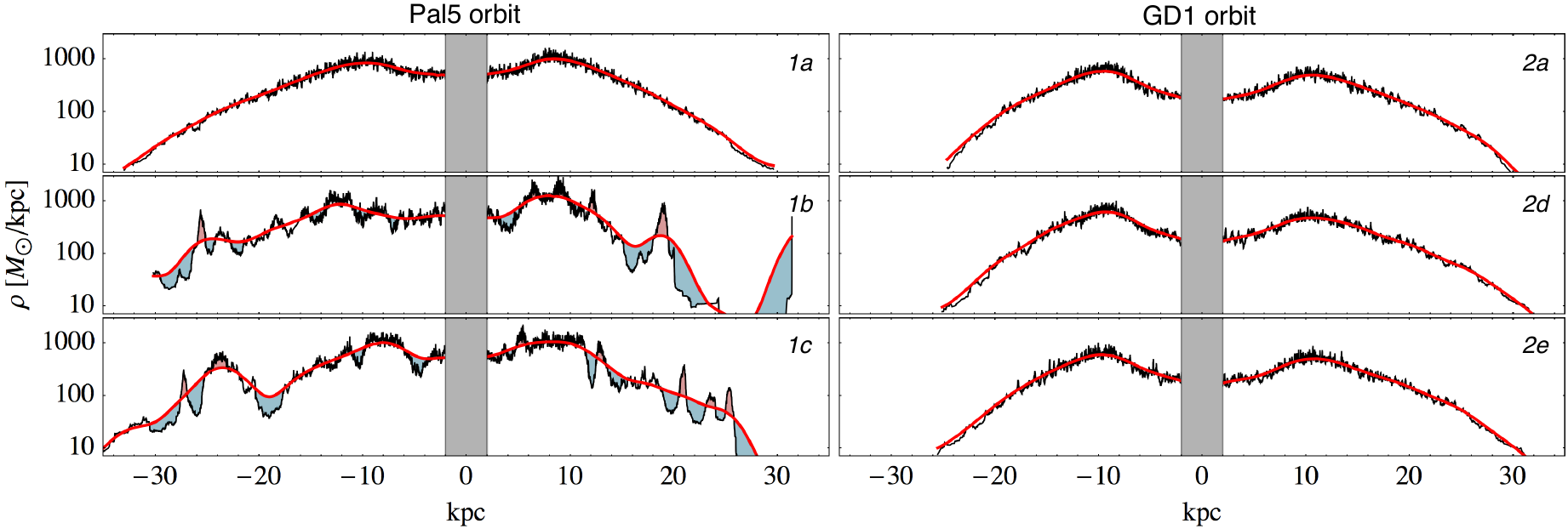}
\caption{Examples of density gaps and clumps; panels are labelled as
  in Fig.~1. In all panels, the black profile shows the 1-d density
  distribution of the stream, while a smoothed profile is shown in
  red.  Shaded gaps and clumps satisfy the selection criteria given in
  equ.~(\ref{sel-cr}). }
\label{fig3}
\end{figure*}

\subsection{Example streams}

Figure 1 displays a few examples of streams from our suite of
simulations of Pal5- and GD1-like orbits (column 1 at $t = 6$ Gyr,
and column 2 at $t = 7$ Gyr). All panels show the two-dimensional
density distribution of the stream when projected onto the plane
perpendicular to the instantaneous orbital angular momentum vector of
the remnant cluster. Streams are straightened using a curvilinear
coordinate system defined by the GC orbit: the $x$ coordinate denotes
distance along the orbit relative to the current position of the
remnant, while the $y$ coordinate denotes distance perpendicular to
the orbit in projection onto this plane.  The remnant itself has been
excised from the plots for convenience. The Galactic centre is in the
negative $y$ direction, so that the orbital motion in all panels is
from left to right.

The top panels, {\it 1a} and {\it 2a}, illustrate runs in which the
streams orbit undisturbed within the smooth Galactic potential,
resulting in unperturbed density distributions. In both columns the
remnant is close to apocenter at the chosen time, so that the
`feathers' are particularly evident, in the form of short diagonal
over-densities located within a few kpc of the remnant.  These are
armlets shed during recent pericentre passages. Their shape is a
consequence of the orbital properties of the member stars
\citep[e.g.,][]{AK12, MB12, NA15}, and is unrelated to external
disturbances.  Panels {\it 1b-c} and {\it 2b-c} show examples of
prograde streams evolved in the presence of a disk of GMCs, each
generated with a different random seed. In all cases, clear
disturbances are evident, in the form of gaps and clumps which are
visually similar to those generated by flybys of DSs \citep[see for
 example Fig. 6 of][]{Ng14}.  { Panels {\it 1d-e} }and {\it 2d-e} show
examples of streams of GCs on retrograde orbits. With respect to the
prograde cases, the relative velocities of encounters are increased
here, reducing their perturbing effects.  

\section{Results}



Figure~2 shows 1-d density profiles for a selection of the streams
displayed in Fig.~1, obtained by collapsing the 2-d distributions
along the $y$ direction. In each column, the top panel corresponds to
the unperturbed stream, while the lower 2 panels show the prograde
cases for Pal5 and the retrograde cases for GD1 (corresponding to the
actual observed orbits). The shaded area $|x|\leq 2$~kpc is excised
from all the analyses that follow, so to avoid possible substructure
due to the feathers. 
{ We isolate any pronounced gaps and clumps and
characterise their abundance and properties.}

Black density profiles are obtained by using a running bin with
adaptive size, containing a fixed number of stream particles, $n=50$.
The bin is moved across the stream one particle at a time. Such
profiles are then smoothed with a Gaussian kernel, to obtain the red
density profiles, of rms width $\lambda$, taken to be a fraction
of the total length of unperturbed streams at that time (the values used are
collected in Table~1). We define gap
(and clump) candidates as all those instances in which the unsmoothed
1-d density profile is continuously below (above) the smoothed
profile, and for each we measure the linear size $l$ and relative mass
contrast $\delta$, defined as
\begin{equation}
    \delta \equiv
    \begin{cases}
    M_{\rm sm}/M& {\rm if}\ M_{\rm sm}\geq M\\
    M/M_{\rm sm}& {\rm if}\ M_{\rm sm}< M
     \end{cases} \ ,
\end{equation}
where $M$ is the mass in the gap or clump, and $M_{\rm sm}$ is the
mass in the same region according to the smoothed density profile. As
shown in Fig.~2, density perturbations show a range of sizes and
mass contrasts. We concentrate on the strongest perturbations, which
are the most interesting for comparison with observation, and which
we define as those satisfying the criteria
\begin{equation}
    \begin{cases}
    \delta> 1.4~;\ l> 0.4\ {\rm kpc}& {\rm for\ gaps,}\\
    \delta> 1.4~;\ l> 0.2\ {\rm kpc}& {\rm for\ clumps.}
     \end{cases} 
     \label{sel-cr}
\end{equation}
No such prominent perturbations are found in our undisturbed
streams. All gaps and clumps satisfying these criteria are shaded in
Fig.~2. { No such structure is found in the two GD1 
streams shown, which only display minor perturbations.}

\begin{figure}
\centering
\includegraphics[width=\columnwidth]{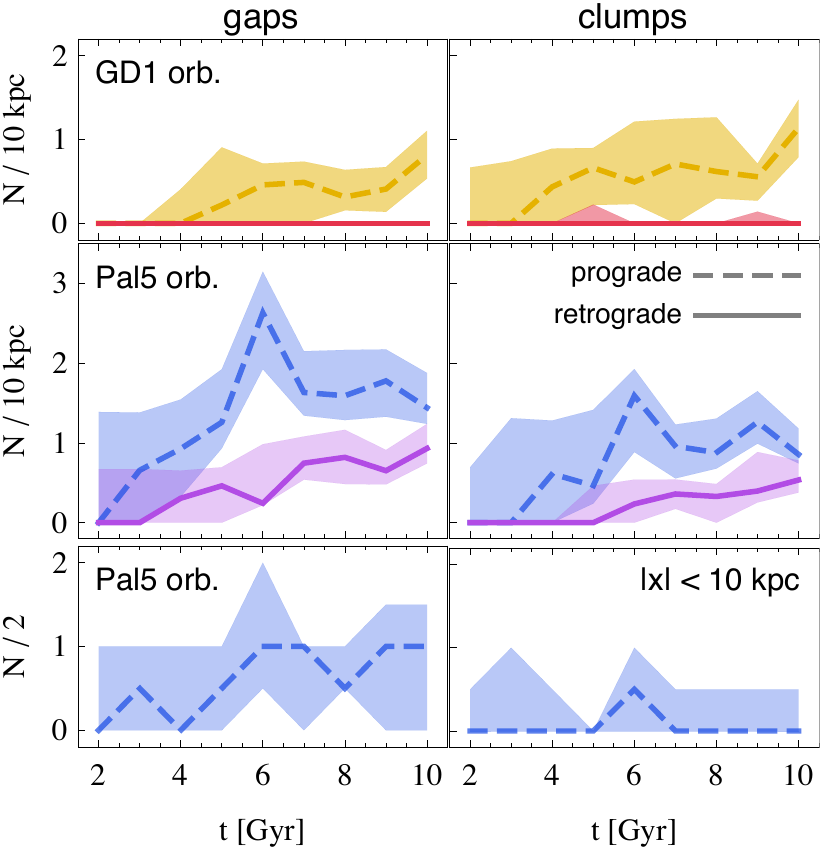}
\caption{Abundances of gaps (left) and clumps (right) satisfying the
  criteria of equ.~$(\ref{sel-cr})$. In all panels, lines show
  the medians, and shaded regions the 10\% to 90\% range, as a
  function of stream age. Top and middle panels show distributions of
  the mean number of perturbations per 10~kpc among the 15 simulations
  carried out for each GC orbit. $N$ in the bottom panels is the
  number of perturbations within 10~kpc of the Pal5 remnant, so that
  $N/2$ estimates the gap (clump) density for the ``observed'' section
  of each simulated Pal5 stream. }
\end{figure}

\subsection{Abundances and properties}

We count all clumps and all gaps that satisfy the criteria of
equ.~(\ref{sel-cr}) in our suite of streams. We have carried out 15
different runs for each of our four GC orbits, and we study how the
abundance of substructure evolves with time, by considering snapshots
at intervals of 1 Gyr between 2 and 10 Gyr.

We first measure the average number of gaps (or clumps) in each
simulation per 10 kpc of stream length, $\langle N/10$~kpc$\rangle$.
We obtain this by dividing the number of perturbations in the stream
by its total length, $L$, defined as the interval in $x$ containing
95\% of the stream particles, $n_{\rm st}$ (excluding those in the excised region $
|x|\leq 2$~kpc; Table~1 lists mean values of $L$ and $n_{\rm st}$ for our streams). Results
are displayed in the upper two rows of panels in Fig.~3: the top
panels pertain to GD1-like orbits, the middle panels to Pal5-like
orbits. Full lines show the medians for retrograde orbits, dashed lines for
prograde orbits.  Shaded regions span the 10\% to 90\% range in each
set of 15 simulations.  A comparison between the top and middle rows
shows that GD1 streams form significantly fewer prominent gaps and
clumps than Pal5 streams. This is a result of the larger
pericentric radius: { the Pal5 stream crosses the disk both  
outside $R_{\odot}$ and within it, where massive GMCs are more abundant. 
The GD1 stream only experiences encounters with 
the GMC population at $R\gtrsim12$~kpc, where the latter has a
substantially lower density and a steeper mass function}. 
For both Pal5- and GD1-like orbits, the lower
encounter velocities of prograde streams result in a systematically
higher density of strong perturbations. The density of such
perturbations increases weakly with time, as more gaps and clumps grow
to exceed the thresholds of equ.~(\ref{sel-cr}).  

\begin{table}
 \centering
 \begin{minipage}{80mm}
  \caption{Mean stream length $L$, mean number of stream particles $n_{\rm st}$ and smoothing length 
  ($\lambda = L_{\rm unp}/30$,  $L_{\rm unp}$ is the length of unperturbed streams) for our 
  prograde Pal5 and retrograde GD1 streams, as a function of time.}
  \begin{tabular}{@{}cccccccccc@{}}
\hline
$t$ $\left[{\rm Gyr}\right]$& 2& 3& 4& 5& 6& 7& 8& 9& 10\\ 
\hline
Pal5&&&&&&&&&\\
$L$ $\left[{\rm kpc}\right]$& 18.1& 19.5& 38.1& 45.8& 46.6& 60.6& 66.6& 82.2& 79.3\\ 
$n_{\rm st}/10^3$& 21.2& 25.3& 31.6& 34.5& 36.2& 41.6& 42.3& 44.0& 47.0\\ 
$\lambda$ $\left[{\rm kpc}\right]$& 0.6& 0.6& 1.2& 1.6& 1.6& 2.0& 2.2& 2.6& 2.6\\ 
\hline
GD1&&&&&&&&&\\
$L$ $\left[{\rm kpc}\right]$& 19.1& 17.7& 27.6& 49.9& 45.9& 47.2& 70.0& 76.7& 77.4\\ 
$n_{\rm st}/10^3$& 11.4& 13.2& 15.9& 18.0& 18.8& 19.8& 21.3& 22.2& 22.6\\ 
$\lambda$ $\left[{\rm kpc}\right]$& 0.6& 0.6& 0.8& 1.7& 1.6& 1.6& 2.3& 2.5& 2.5\\ 
\hline
\end{tabular}
\end{minipage}
\end{table}

It is not currently known whether a bound remnant of GD1 still exists,
nor where it would be located with respect to the observed section of
the stream. Our average perturbation density can thus, perhaps, be
considered representative for this stream.  Fig.~3 shows that its
retrograde motion and large pericentric radius result in very few
perturbations as prominent as required by equ.~(\ref{sel-cr}). 
Recall, however, that our model is conservative in that we ignore 
the mass-size relation of GMCs \citep[e.g.,][]{Ev99,NM11}, and
soften all GMC particles on the same scale, $\epsilon=100$~pc.  
This results in a uniform half-mass radius of $r_{\rm h}\approx120$~pc,
an overestimate for the bulk of the population. 
{Though rare, this is relevant in those cases in which the point 
of closest approach during the flyby is smaller than this scale.}


On the other hand, the average measurements just presented may
overestimate the number of perturbations in the observed section of
the Pal5 stream, which only extends about $10$~kpc from the remnant.
Material in this region has been shed more recently (and so has
experienced fewer disk crossings) than material further down the
tails. In addition, it is dynamically younger so that any
perturbations have had less time to grow above the thresholds of
equ.~(\ref{sel-cr})).  For our Pal5 prograde streams, we therefore
count the number of strong gaps and clumps within 10~kpc of the
remnant, $N$. Results are plotted in the bottom panels of Fig.~3, as
$\langle N/2\rangle$, providing an estimate of the number of perturbations expected
within the observed section of the Pal5 stream. The upper panels of
Figure~4 illustrate the range of properties of such gaps and clumps,
by collecting all of those we identify in the younger section,
$|x|<10$~kpc, of our prograde Pal5-like runs. Color coding indicates
time and the dashed lines identify the criteria~(\ref{sel-cr}).  The
populations of both gaps and clumps extend to sizes of $\sim 2$~kpc
and gaps reach mass contrasts $\delta\gtrsim$ 2.  The lower panels of
the same Figure collect substructure located over the full length of
the prograde Pal5 streams. This increases the range of sizes and mass
contrasts even further. As `older' gaps and clumps that have streamed
far from the remnant are included, a color gradient across the scatter
plots becomes apparent, illustrating how the internal dynamics of the
stream amplifies perturbations \citep[see e.g.,][]{DE15a}.

\begin{figure}
\centering
\includegraphics[width=\columnwidth]{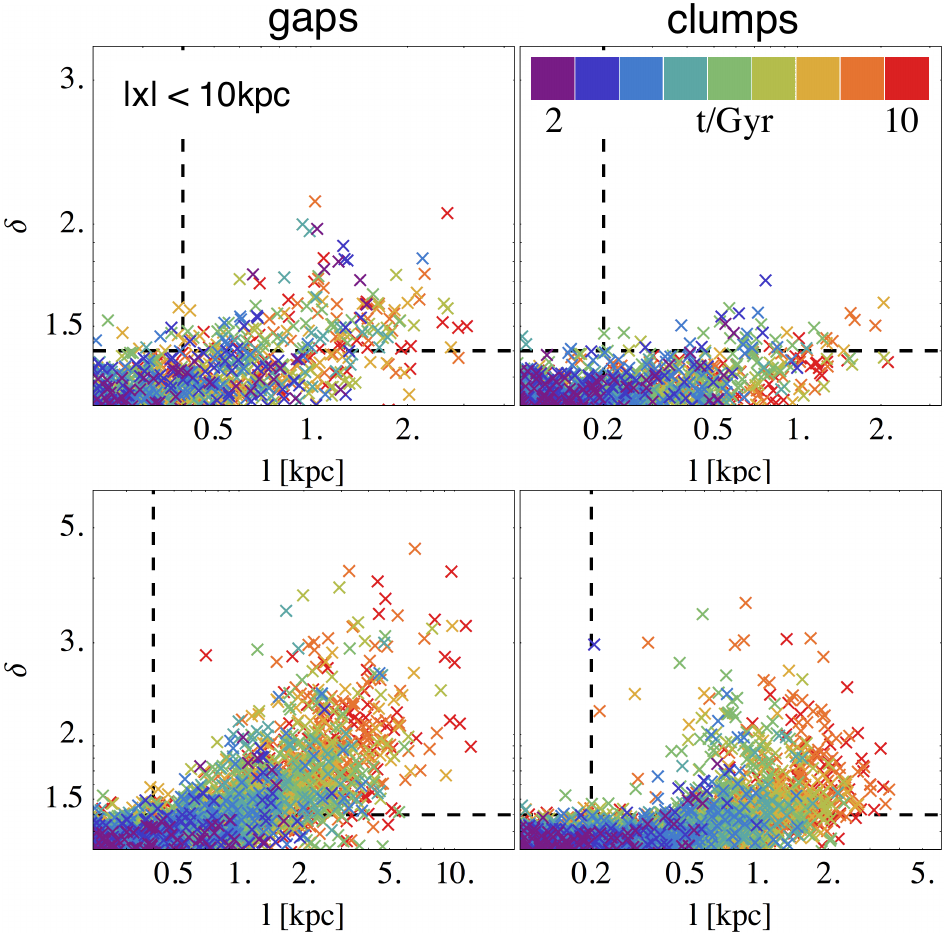}
\caption{Properties of clumps (right) and gaps (left) in our Pal5-like
  streams, as a function of stream age (encoded by colour). The top
  panels show results for the `observed' regions within 10~kpc from
  the remnant, the lower panels for the streams as a whole.}
\label{fig5}
\end{figure}

\section{Discussion}

We have shown that the Galactic population of GMCs can significantly
perturb kinematically cold globular cluster
streams. Using a flat prior (3 to 10 Gyr) for the age of the Pal5
stream we find that the observed $\sim 10$~kpc long section of the
stream is expected to show {$0.5^{+0.5}_{-0.5}$} gaps (median and 10\% and
90\% quantiles; { 0.65$\pm$0.5, mean with one standard deviation}) 
wider than 0.4~kpc and with a mass contrast
$\delta>1.4$. Our GD1 streams, in contrast, experience 
fewer strong GMC perturbations because of their larger pericentre and
{ the steeper mass function of the GMCs in the outer disk \citep{HD15, Ri16}}.  
Additionally, they are less sensitive to each encounter
because of the higher relative velocities resulting from the
retrograde orbit. Structures more prominent than the thresholds of
equ.~(\ref{sel-cr}) are rare in our GD1 streams, { at least in our 
conservative numerical setup, in which all GMCs are smoothed on a 
scale of $\epsilon=100$~pc.} Weaker
but significant gaps and clumps are still present, and may be
detectable as enhanced noise in sufficiently deep observations. We
defer the quantification of such perturbations to follow-up work.

Currently available observations do not seem to show obvious evidence
for prominent gaps \citep[e.g., ][]{RI16,RC16,GT16}.  However, our results
suggest that deep enough observations of the Pal5 stream (and possibly
of the GD1 stream) should detect perturbations of GMC origin, even if
there is no structure induced by dark matter subhaloes.  This re-sets
expectations for GC streams in Milky Way models with smooth haloes.
\citet{RC12,RC16} find the Pal5 and GD1 streams to display more
substructure than is expected from counting noise alone.  From a
statistical standpoint, more detailed studies are needed both of
observational systematics such as photometric and
background-subtraction uncertainties and of theoretical systematics
reflecting uncertainties in the stream structure expected in the
absence of small-scale perturbations. Only then will it be possible to
establish whether apparent inhomogeneities {such as those 
marginally detected in the Pal5 stream by \citet{JB16}} result from encounters with
GMCs or with DSs. Distinguishing between these two is made challenging by
the absence of clear predictions for the abundance of DM substructure
in the haloes of disk galaxies like the MW, where disk shocking and
the enhanced tidal field destroy DSs more easily than in DM-only
simulations \citep[][]{ED10,JP10}. This is particularly uncertain at
the small galactocentric distances of observed GC streams.

In the event of a clear detection of density perturbations of external
origin, direct modelling may allow DS and GMC perturbers to be
distinguished. \citet{DE15b} have shown that by using a combination of
both photometric and kinematic data it may be feasible to infer the
time since flyby, and the approximate location of the encounter.
Given their general framework, the highly constrained spatial and
relative velocity distributions of GMC encounters may allow encounters
that took place with objects on prograde circular orbits in the disk plane
to be distinguished from those that occurred well away from the plane
and at very different relative velocities.

\section*{Acknowledgments}
NA and SV thank Denis Erkal for stimulating discussions. {We thank the referee, Ray Carlberg, for a constructive report.}


\end{document}